\documentclass[aps,10pt,twocolumn,prb,amsmath,amssymb,showpacs,superscriptaddress]{revtex4-1}
\usepackage{graphicx}
\pdfoutput=1

\begin{document}
\title{Universal quantum computation with ordered spin-chain networks}
\author{Yaroslav Tserkovnyak}
\affiliation{Department of Physics and Astronomy, University of California, Los Angeles, California 90095, USA}
\author{Daniel Loss}
\affiliation{Department of Physics, University of Basel, Klingelbergstrasse 82, CH-4056 Basel, Switzerland}
 
\begin{abstract}
It is shown that anisotropic spin chains with gapped bulk excitations and magnetically ordered ground states offer a promising platform for quantum computation, which bridges the conventional single-spin-based qubit concept with recently developed topological Majorana-based proposals. We show how to realize the single-qubit Hadamard, phase, and $\pi/8$ gates as well as the two-qubit CNOT gate, which together form a fault-tolerant universal set of quantum gates. The gates are implemented by judiciously controlling Ising exchange and magnetic fields along a network of spin chains, with each individual qubit furnished by a spin-chain segment. A subset of single-qubit operations is geometric in nature, relying on control of anisotropy of spin interactions rather than their strength. We contrast topological aspects of the anisotropic spin-chain networks to those of $p$-wave superconducting wires discussed in the literature.
\end{abstract}



\maketitle

\section{Introduction}

Quantum control of electronic systems is usually restricted to the realm of very small structures, wherein qubits are engendered by elementary quantum degrees of freedom like electron spin (or, more generally, Kramers doublet). Notable exceptions are provided by macroscopically coherent superconducting qubits\cite{[{}][{, and references therein. }]makhlinRMP01} and anyons with non-Abelian braiding statistics.\cite{[{}][{, and references therein. }]nayakRMP08} Such ``nonabelions" may arise in certain correlated topological states of matter\cite{mooreNPB91} (most prominently, the $\nu=5/2$ fractional quantum Hall effect) or be fabricated in topological-insulator based systems\cite{fuPRL08} and even rather conventional semiconductor-based heterostructures.\cite{sauPRL10} The latter require tailoring an appropriate combination of spin-orbit interactions, magnetism (or magnetic field), and superconductivity in order to furnish a topological state with non-Abelian Majorana fermions. The localized and spatially separated Majorana fermions can endow the ground state with necessary degeneracy\cite{ivanovPRL01} and robust quantum coherence,\cite{kitaevUFN01,*kitaevCHA10} which in turn provide a basis for topologically-protected quantum computation.\cite{aliceaNATP11}

Despite a promise for solid-state quantum computation, both the superconductor-based and topological (almost always Majorana-based) platforms are not without serious challenges, however. The former suffers from relatively fast (while poorly understood) decoherence, which is perhaps related to its macroscopic nature and nonsuperfluid parasitic low-lying degrees of freedom, while the latter has in general only a subset of quantum gates that is topologically protected, which needs to be supplemented by more traditional schemes (for example, based on interaction between Majoranas,\cite{bravyiPRA06} coupling to flux qubits,\cite{hasslerNJP10} or an rf measurement\cite{sauPRA10}), in order to realize a universal set of gates.\cite{nielsenBOOK00} Overcoming these obstacles poses tremendous technological challenges.

At the same time, steady progress is being made in harnessing single-electron spins in semiconductor quantum dots as natural building blocks for universal quantum computation.\cite{lossPRA98,*[{}][{, and references therein. }]zakRNC10} While a serious problem is posed by nuclear-spin induced qubit decoherence, record-long coherence times have been achieved,\cite{bluhmCM10lc} which are four orders of magnitude larger than those in early experiments.\cite{pettaSCI05}

In this paper, we are proposing a hybrid platform for universal quantum computation, which is based on anisotropic spin chains. A chain can consist of a finite sequence of several appropriately engineered and tunable quantum dots or a section of magnetically ordered one-dimensional wire. In the absence of magnetic fields, the degeneracy of such a spin chain is guaranteed by time-reversal symmetry. In the presence of weak magnetic fields, the degeneracy can survive up to a critical field, if the spin chain is sufficiently long on the scale of an appropriate coherence length. A nonlocal topological character of spin-chain qubits may be inferred from the sensitivity of the quantum coherence within the ground-state subspace to the boundary conditions. Furthermore, a formal similarity\cite{kitaevUFN01} of the nonlocal spin qubits in open chains with Majorana end states in topological superconducting wires is revealed by a Jordan-Wigner transformation,\cite{liebANP61} allowing one to tap into the existing pool of ideas on Majorana-based quantum computation (while appreciating fundamental differences between spin- and superconductor-based networks of such wires, as discussed in the Appendix).

We will construct geometric gates on our spin systems that mimic braiding of nonabelions, while the readout can be performed via standard spin-to-charge conversion schemes\cite{lossPRA98} on individual sites or dots. The readout can also be accomplished by fusing the spin-chain qubits and measuring spin torque, in analogy to the Josephson current readout of Ref.~\onlinecite{aliceaNATP11}. While drawing inspiration from both worlds of (spin-based and topological) quantum computation, we are motivated by a versatility for practical realizations of such hybrid systems as much as bridging and synthesizing diverse theoretical ideas that have been recently developed in different subareas of quantum computation.

\section{Single-qubit gates}

The nonlocal spin qubits arising in anisotropic spin chains are defined (along with a review of necessary introductory material) in the Appendix. As a key result of this section, we will construct a phase gate on such qubits, which is analogous to braiding of two vortices in chiral $p$-wave superconductors\cite{ivanovPRL01} or two Majoranas in superconducting nanowire networks.\cite{aliceaNATP11} While our phase gate will be geometric in spirit, no braiding is necessary, however. In fact, braiding of our Majoranas as in Ref.~\onlinecite{aliceaNATP11} would not generate a nontrivial gate.\cite{Note1} (See also Ref.~\onlinecite{flensbergCM10na} for braiding-less gates on Majoranas in superconducting wires coupled to single-electron quantum dots.)

Before proceeding, let us specify our single-spin basis:
\begin{equation}
|\theta,\phi\rangle=e^{i\phi/2}e^{-i\phi\hat{\sigma}_z/2}e^{-i\theta\hat{\sigma}_x/2}|{\uparrow}\rangle=\cos\frac{\theta}{2}|{\uparrow}\rangle+e^{i\phi}\sin\frac{\theta}{2}|{\downarrow}\rangle\,,
\end{equation}
where $(\theta,\phi)$ are polar angles (over the Bloch sphere), which define spin direction of a coherent-spin state, and $|{\uparrow(\downarrow)}\rangle=\delta_{s,\uparrow(\downarrow)}$ are the usual spin-$1/2$ up or down eigenstate along the $z$ axis. The singularity (Dirac string) of our basis is along the south pole on the Bloch sphere. The corresponding Berry phase\cite{berryPRSLA84} is given by
\begin{equation}
\delta\varphi_B=-i\langle\theta,\phi|\delta|\theta,\phi\rangle=\frac{1}{2}(1-\cos\theta)\delta\phi\,.
\label{fB}
\end{equation}

In this section, we will construct single-qubit gates for an Ising chain. To provide an explicit demonstration, we start with a three-site segment [which  is readily generalizable to longer chains, as sketched in Fig.~\ref{fig2}(b) below; in practice, however, we would be interested in a mesoscopic situation, such that the chains are long enough on the scale of the appropriate correlation length $\xi$, to have the degenerate ground state, but not too long to avoid inevitable decoherence effects and errors associated with the control of magnetic anisotropies] described by the following time-dependent Hamiltonian:
\begin{equation}
\frac{H}{J}=-\hat{\sigma}^x_1\hat{\sigma}^x_2-\left[\boldsymbol{\hat{\sigma}}_2\cdot\mathbf{m}(t)\right]\left[\boldsymbol{\hat{\sigma}}_3\cdot\mathbf{m}(t)\right]-\sum_{i=1}^3h_i(t)\hat{\sigma}^z_i\,.
\label{123}
\end{equation}
When $\mathbf{m}=\mathbf{x}$ and $h_i=0$, the chain is in some superposition of the ground states $|{\Rightarrow}\rangle$ and $|{\Leftarrow}\rangle$, Eq.~(\ref{gs}).

We first discuss a straightforward way to initialize the system in a Greenberger-Horne-Zeilinger (GHZ) state $|{+}\rangle$, where [see also Eq.~(\ref{01})]
\begin{equation}
|\pm\rangle=\frac{|{\Rightarrow}\rangle\pm|{\Leftarrow}\rangle}{\sqrt{2}}
\label{ls}
\end{equation}
are the logic states for three spins (and analogously for $N$ spins). Ramping the uniform field $h_i\to h$ in the chain with $\mathbf{m}=\mathbf{x}$ up to a large positive value establishes the unique ground state $|{\Uparrow}\rangle$. Slowly then turning off $h_1(t)$ to zero leaves the first site in the $|{\uparrow}\rangle=(|{\rightarrow}\rangle+|{\leftarrow}\rangle)/\sqrt{2}$ state, while the second and third sites remain pinned in the same up state by the magnetic field along the $z$ axis. Assuming, that $h$ is sufficiently large, we can neglect quantum fluctuations of the pinned spins, so that, since $\langle\hat{\sigma}^x_2\rangle=0$, the first site is completely decoupled from the rest of the chain. Turning then $h_2(t)$ slowly off, the second site picks coherently the orientation of the first site along the $x$ axis, while the last site still remains decoupled by the pinning field. This gives the following state for the first two spins: $(|{\rightarrow\rightarrow}\rangle+|{\leftarrow\leftarrow}\rangle)/\sqrt{2}$. In order to see this explicitly, we write the initial state for the first two spins as
\begin{equation}
\psi=\frac{|{\rightarrow}\rangle+|{\leftarrow}\rangle}{\sqrt{2}}\otimes|{s}\rangle\,,
\end{equation}
where $|{s}\rangle$ describes the state of the second spin (effectively decoupled from the third spin). Since the interaction between the two spins is of the $x$-Ising form, which cannot flip the first spin along the $x$ axis, their time-dependent state will have the form
\begin{equation}
\psi(t)=\frac{|{\rightarrow}\rangle\otimes|{s}_+(t)\rangle+|{\leftarrow}\rangle\otimes|{s}_-(t)\rangle}{\sqrt{2}}\,,
\label{psi12}
\end{equation}
with the initial condition $|{s}_+(t)\rangle=|{s}_-(t)\rangle=|{s}\rangle$. Direct substitution of Eq.~(\ref{psi12}) into the Schr{\"o}dinger equation then immediately shows that the second spin $|{s}_\pm(t)\rangle$ evolves adiabatically according to the effective magnetic field $\mathbf{h}_{\rm eff}=\pm J\mathbf{x}+h_2(t)\mathbf{z}$, thus ending up in the entangled $(|{\rightarrow\rightarrow}\rangle+|{\leftarrow\leftarrow}\rangle)/\sqrt{2}$ state. Note that in this process, Berry phases (\ref{fB}) vanish in our basis, since $\delta\phi=0$.

\begin{figure}
\includegraphics[width=\linewidth]{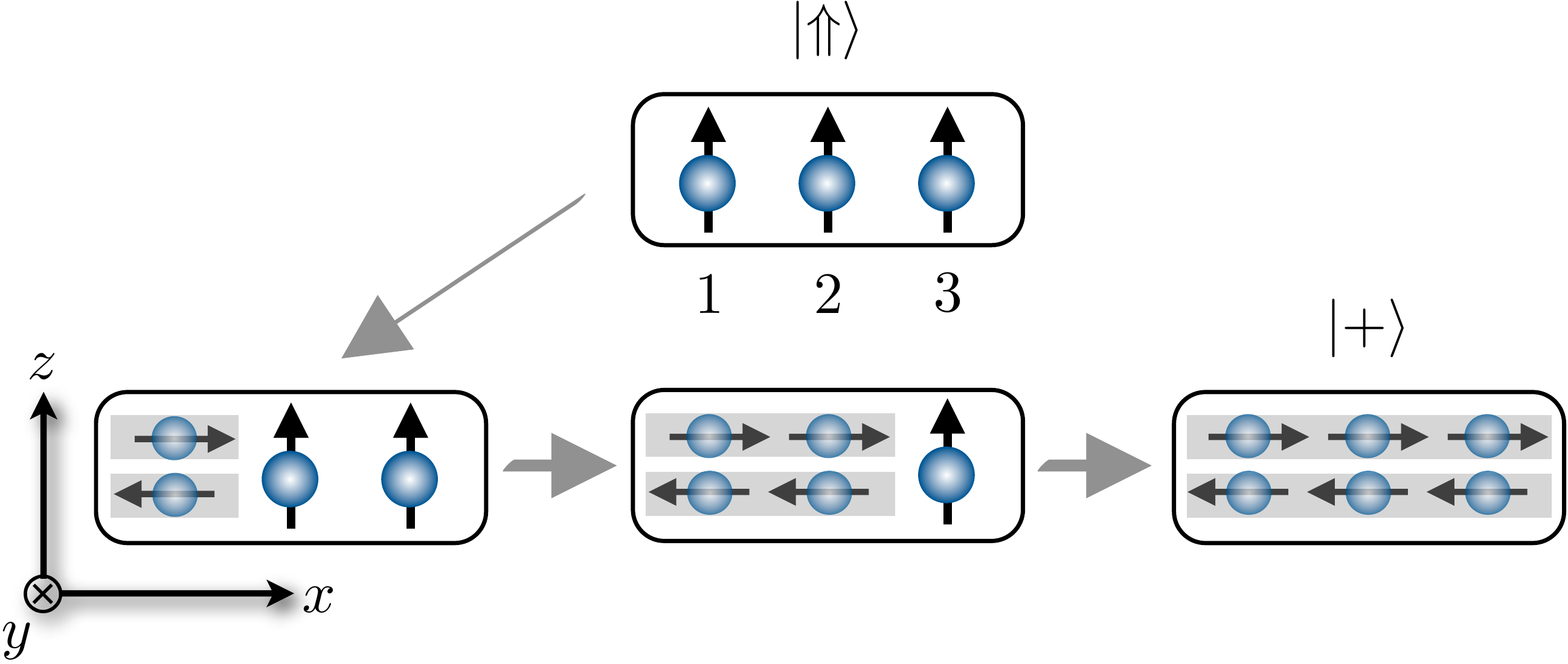}
\caption{(color online). Initialization of an $N$-site ($N=3$ as drawn) chain in a Greenberger-Horne-Zeilinger state $|{+}\rangle$. See text for details.}
\label{fig1}
\end{figure}

\begin{figure}
\includegraphics[width=\linewidth]{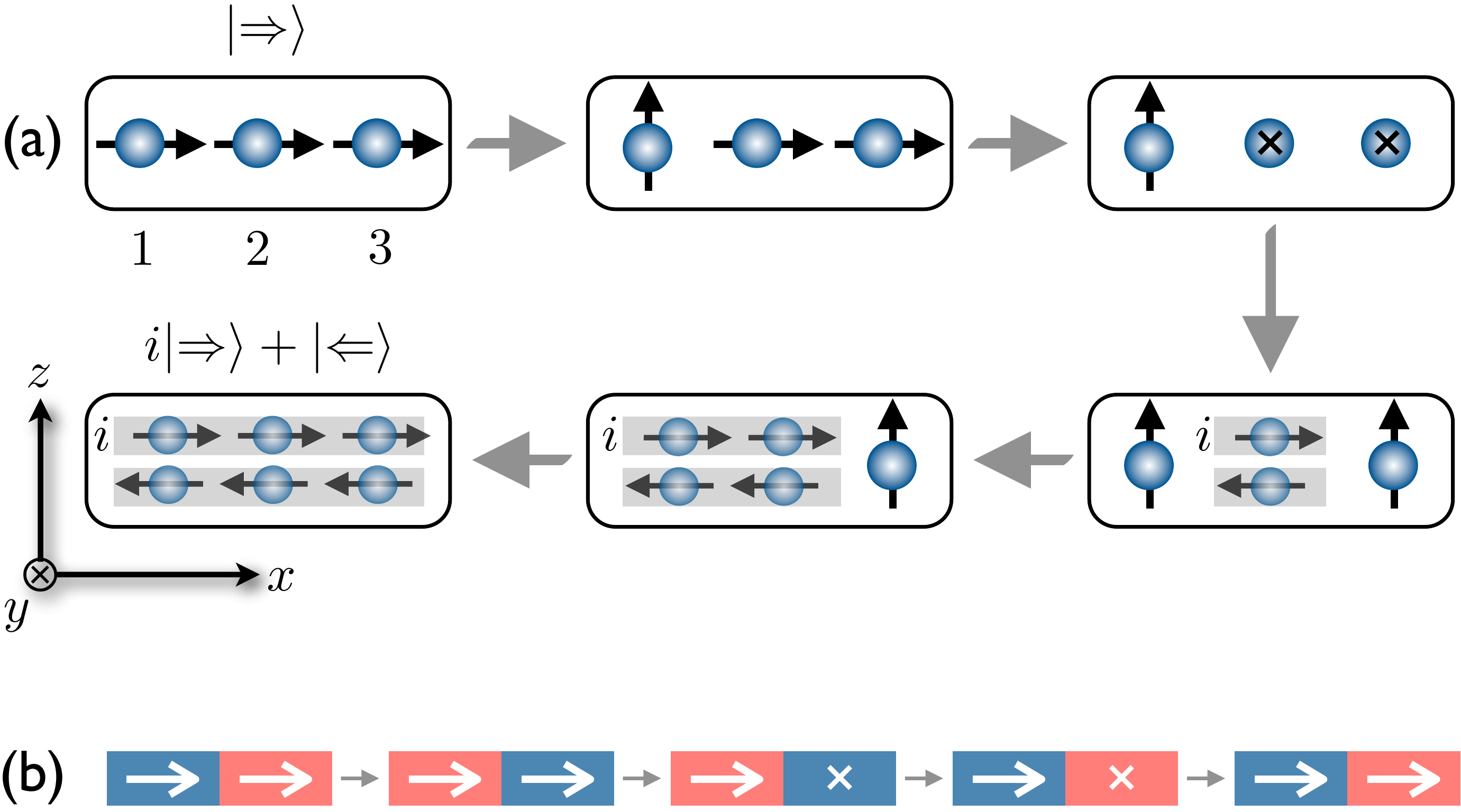}
\caption{(a) Application of the phase gate in the $|{\pm}\rangle$ basis to a three-site spin chain $|{\Rightarrow}\rangle$, as explained in the text. (b) Crude schematics for a longer chain: The blue regions show the ``topological" Ising sections in the absence of magnetic field, with degenerate ground states. The red regions are ordinary nondegenerate spin-polarized sections subjected to a large magnetic field in the $z$ direction. The white arrows show the direction of the Ising interaction axis at each stage. The topological region encoding a qubit is first pushed to the right by raising the magnetic field to its left and lowering it to its right. The key operation is rotating the Ising axis from the $x$ direction to $y$ in the second step. The Hamiltonian returns to its original form in the last step, while the qubit undergoes the phase-gate transformation (\ref{pm}). Note that there need not be a full control of individual sites, only anisotropies over the relevant segments of the entire chain.}
\label{fig2}
\end{figure}

Finally, turning off $h_3(t)$, we establish the GHZ ground state $|{+}\rangle$, Eq.~(\ref{ls}). The schematic of this procedure, which ``zips" the state $|{\Uparrow}\rangle$ into $|{+}\rangle$ by gradually turning off magnetic field on sites 1, 2, etc., is shown in Fig.~\ref{fig1}. Any initial quantum coherence in the logic space would be wiped out by this initialization. It is, furthermore, clear that this initialization works for an arbitrary number of spins in the chain.\cite{Note2}

We now implement the phase gate
\begin{equation}
\left(\begin{array}{c}+ \\ -\end{array}\right)\to\left(\begin{array}{cc}1 & 0 \\ 0 & i\end{array}\right)\left(\begin{array}{c}+ \\ -\end{array}\right)
\label{pm}
\end{equation}
on this three-site system as follows [see Fig.~\ref{fig2}(a) for a schematic]. Let us start in the $|{\Rightarrow}\rangle$ state with $\mathbf{m}=\mathbf{x}$ and $h_i=0$ in Eq.~(\ref{123}). We first ramp field $h_1$ up to a large positive value, which tilts the first spin in the $xz$ plane from the $x$ toward the $z$ direction, contributing a vanishing Berry phase (\ref{fB}) as $\delta\phi=0$. Then $\mathbf{m}$ is adiabatically rotated in the $xy$ plane from $\mathbf{x}$ to $\mathbf{y}$. This step rotates the remaining two spins in the $xy$ plane from the $x$ direction to the $y$ direction, which contributes the Berry phase $\pi/4$ per spin, since $\cos\theta=0$ and $\delta\phi=\pi/2$. The third spin is then biased by a large positive field $h_3$ (no Berry phase). The second spin is now effectively decoupled and in the state
\begin{equation}
|{\pi/2,\pi/2}\rangle=\frac{|{\uparrow}\rangle+i|{\downarrow}\rangle}{\sqrt{2}}=\frac{e^{i\pi/4}|{\rightarrow}\rangle+e^{-i\pi/4}|{\leftarrow}\rangle}{\sqrt{2}}\,.
\end{equation}
When the field $h_1$ is slowly turned off at the next step, we thus arrive at the state
\begin{equation}
\frac{e^{i\pi/4}|{\rightarrow\rightarrow}\rangle+e^{-i\pi/4}|{\leftarrow\leftarrow}\rangle}{\sqrt{2}}
\end{equation}
for the first two spins. Rotating $\mathbf{m}$ back to $x$ and turning off $h_3$ returns us to the original Hamiltonian, while the wave function transforms to the state
\begin{equation}
|{\Rightarrow}\rangle\to\frac{e^{i\pi/4}|{\Rightarrow}\rangle+e^{-i\pi/4}|{\Leftarrow}\rangle}{\sqrt{2}}\,,
\label{RR}
\end{equation}
neglecting the aforementioned Berry phase, as well as the dynamic phase, which are the same for the other initial state $|{\Leftarrow}\rangle$. Repeating these steps for the $|{\Leftarrow}\rangle$ initial state, we obtain the transformation:
\begin{equation}
|{\Leftarrow}\rangle\to\frac{e^{-i\pi/4}|{\Rightarrow}\rangle+e^{i\pi/4}|{\Leftarrow}\rangle}{\sqrt{2}}\,.
\label{LL}
\end{equation}
Putting Eqs.~(\ref{RR}) and (\ref{LL}) together, we get precisely the phase gate (\ref{pm}) in the logic-space basis (\ref{ls}). The extension of this scheme to longer chains is straightforward, which is crudely sketched in Fig.~\ref{fig2}(b). We note in passing that this geometric procedure is analogous to Majorana braiding in superconducting wire $T$ junctions,\cite{aliceaNATP11} which effectively flips the superconducting phase by $\pi$. In our spin chain, rotating Ising spins by $\pi/2$ in the $xy$ plane is in fact fully analogous to changing the superconducting phase by $\pi$ for the Jordan-Wigner fermions (see Sec.~\ref{fuse}).

Partial rotations of the Ising interaction axis from $\mathbf{x}$ to some direction $\mathbf{m}$ in the $xy$ plane, at the intermediate step of the above procedure, could produce an arbitrary desired rotation around the $z$ axis on the Bloch sphere of the logic space. Namely, for $\mathbf{m}=(\cos\phi,\sin\phi,0)$, we find
\begin{equation}
\left(\begin{array}{c}+ \\ -\end{array}\right)\to\left(\begin{array}{cc}1 & 0 \\ 0 & e^{i\phi}\end{array}\right)\left(\begin{array}{c}+ \\ -\end{array}\right)
\label{phi}
\end{equation}
The case $\phi=\pi/4$, in particular, realizes the $\pi/8$ gate, which in the proposal of Ref.~\onlinecite{sauPRA10} would require bringing Majoranas together and lifting the degeneracy in the logic-space basis.

Finally, a Hadamard-like gate (or likewise an arbitrary rotation around the $x$ axis on the Bloch sphere of the logic space) is easily realized by applying a magnetic field pulse $\varphi=Jh\Delta t$ along the $x$ axis to, say, the first spin of the chain, so that
\begin{equation}
\left(\begin{array}{c}\Rightarrow \\ \Leftarrow\end{array}\right)\to\frac{1}{\sqrt{2}}\left(\begin{array}{cc}e^{i\varphi} & 0 \\ 0 & e^{-i\varphi}\end{array}\right)\left(\begin{array}{c}\Rightarrow \\ \Leftarrow\end{array}\right)\,.
\label{RL}
\end{equation}
For $\varphi=\pi/4$, this gives
\begin{equation}
\left(\begin{array}{c}+ \\ -\end{array}\right)\to\frac{1}{\sqrt{2}}\left(\begin{array}{cc}1 & i \\ i & 1\end{array}\right)\left(\begin{array}{c}+ \\ -\end{array}\right)\,.
\label{Hl}
\end{equation}
The Hadamard gate can be then obtained from this with the help of two phase gates:
\begin{equation}
\left(\begin{array}{cc}1 & 1 \\ 1 & -1\end{array}\right)=\left(\begin{array}{cc}1 & 0 \\ 0 & -i\end{array}\right)\left(\begin{array}{cc}1 & i \\ i & 1\end{array}\right)\left(\begin{array}{cc}1 & 0 \\ 0 & -i\end{array}\right)\,.
\end{equation}

Notice a certain symmetry between the bases (i) $\{|{+}\rangle,|{-}\rangle\}$ and (ii) $\{|{\Rightarrow}\rangle,|{\Leftarrow}\rangle\}$. In the basis (i), we can implement an arbitrary phase gate (\ref{phi}) by rotating the Ising interaction axis $\mathbf{m}$, as described above, and a Hadamar-like gate (\ref{Hl}) by the magnetic field pulse $\varphi=\pi/4$. In basis (ii), an arbitrary phase gate (\ref{RL}) is realized by a magnetic-field pulse, while the Hadamar-like gate, Eqs.~(\ref{RR}) and (\ref{LL}), by the $\pi/2$ rotation of the Ising interaction axis in the $xy$ plane. According to the preceding discussion, we can thus realize the desired set of the pure Hadamar, phase, and $\pi/8$ gates in either of the two bases.

\section{Two-qubit gates}

For the two-qubit gates, consider two Ising spin chains put in parallel, each described by Hamiltonian (\ref{Hi}) with $h=0$. These qubits can be implemented as mobile topological sections embedded in longer spin chains and separated by nontopological regions as sketched in the example of Fig.~\ref{fig2}(b). We can implement a two-qubit operation by pulsing an $x$-Ising exchange interaction $J$ between, say, the first sites in each qubit. In the basis $\{|{\Rightarrow}\rangle,|{\Leftarrow}\rangle\}$ for each qubit, the transformation amounts to
\begin{align}
|{\Rightarrow\Rightarrow}\rangle\to e^{i\phi}|{\Rightarrow\Rightarrow}\rangle&\,,\,\,\,|{\Leftarrow\Leftarrow}\rangle\to e^{i\phi}|{\Leftarrow\Leftarrow}\rangle\,,\nonumber\\
|{\Rightarrow\Leftarrow}\rangle\to e^{-i\phi}|{\Rightarrow\Leftarrow}\rangle&\,,\,\,\,|{\Leftarrow\Rightarrow}\rangle\to e^{-i\phi}|{\Leftarrow\Rightarrow}\rangle\,,
\label{RRLL}
\end{align}
where $\phi=J\Delta t$. For the special case of $\phi=-\pi/4$, Eqs.~(\ref{RRLL}) result in the following gate in the logic-space basis (\ref{ls}):
\begin{equation}
\left(\begin{array}{c}++ \\ -+ \\ +- \\ --\end{array}\right)\to\frac{1}{\sqrt{2}}\left(\begin{array}{cccc}1 & 0 & 0 & -i \\ 0 & 1 & -i & 0 \\ 0 & -i & 1 & 0 \\ -i & 0 & 0 & 1\end{array}\right)\left(\begin{array}{c}++ \\ -+ \\ +- \\ --\end{array}\right)\,.
\label{ppmm}
\end{equation}

At this point, it is instructive to make a brief digression and recall the braid-group representation of four vortices in chiral $p$-wave superconductors.\cite{ivanovPRL01} The vortices define four Majoranas, $\{\gamma_1,\gamma_2,\gamma_3,\gamma_4\}$, which can be combined into two complex fermions $c_1=(\gamma_1+i\gamma_2)/2$ and $c_2=(\gamma_3+i\gamma_4)/2$. Braiding on a plane is accomplished by three generators, $\{T_{12},T_{23},T_{34}\}$, respectively exchanging vortices 1 and 2, 2 and 3, and 3 and 4. These generators are represented by
\begin{equation}
\tau_{ij}=\tau(T_{ij})=e^{\pi\gamma_j\gamma_i/4}=\frac{1+\gamma_j\gamma_i}{\sqrt{2}}\,,
\label{Tij}
\end{equation}
which introduces a relative sign between two Majoranas upon their exchange:
\begin{equation}
\gamma_i\to\tau_{ij}\gamma_i\tau^\dagger_{ij}\,,
\end{equation}
such that $\gamma_i\to\gamma_j$ and $\gamma_j\to-\gamma_i$ upon braiding vortices $i\neq j$.\cite{Note3} Written in the basis $\{1,c_1^\dagger,c_2^\dagger,c_1^\dagger c_2^\dagger\}|{0}\rangle$, the braid group representation (\ref{Tij}) is given explicitly by\cite{ivanovPRL01}
\begin{widetext}
\begin{equation}
\tau_{12}=\left(\begin{array}{cccc}e^{-i\pi/4} & 0 & 0 & 0 \\ 0 & e^{i\pi/4} & 0 & 0 \\ 0 & 0 & e^{-i\pi/4} & 0 \\ 0 & 0 & 0 & e^{i\pi/4}\end{array}\right)\,,\,\,\,
\tau_{23}=\frac{1}{\sqrt{2}}\left(\begin{array}{cccc}1 & 0 & 0 & -i \\ 0 & 1 & -i & 0 \\ 0 & -i & 1 & 0 \\ -i & 0 & 0 & 1\end{array}\right)\,,\,\,\,\tau_{34}=\left(\begin{array}{cccc}e^{-i\pi/4} & 0 & 0 & 0 \\ 0 & e^{-i\pi/4} & 0 & 0 \\ 0 & 0 & e^{i\pi/4} & 0 \\ 0 & 0 & 0 & e^{i\pi/4}\end{array}\right)\,.
\label{br}
\end{equation}
\end{widetext}
States $\{1,c_1^\dagger c_2^\dagger\}|{0}\rangle$ and $\{c_1^\dagger,c_2^\dagger\}|{0}\rangle$ have different fermion parities and are thus not mixed by braiding, which conserves the parity.\cite{ivanovPRL01} We notice that $\tau_{23}$ coincides with transformation (\ref{ppmm}) in our two-qubit logic space, while $\tau_{12}$ and $\tau_{34}$ simply reproduce phase gate (\ref{pm}) performed individually on one of the two spin chains.

Since such standard braiding operations (\ref{br}) are not sufficient for a universal quantum computation, we switch the logic-space basis from $\{|{+}\rangle,|{-}\rangle\}$ to $\{|{\Rightarrow}\rangle,|{\Leftarrow}\rangle\}$ and supplement transformation (\ref{RRLL}) ($\phi=-\pi/4$) with single-qubit phase gate (\ref{RL}) ($\varphi=\pi/4$) for each qubit. This gives (up to an overall phase)
\begin{equation}
\left(\begin{array}{c}\Rightarrow\Rightarrow \\ \Leftarrow\Rightarrow \\ \Rightarrow\Leftarrow \\ \Leftarrow\Leftarrow\end{array}\right)\to\frac{1}{\sqrt{2}}\left(\begin{array}{cccc}1 & 0 & 0 & 0 \\ 0 & 1 & 0 & 0 \\ 0 & 0 & 1 & 0 \\ 0 & 0 & 0 & -1\end{array}\right)\left(\begin{array}{c}\Rightarrow\Rightarrow \\ \Leftarrow\Rightarrow \\ \Rightarrow\Leftarrow \\ \Leftarrow\Leftarrow\end{array}\right)\,,
\end{equation}
which is the controlled-$Z$ gate. Sandwiching it by the Hadamard gates applied to the first qubit realizes the CNOT gate
\begin{equation}
\left(\begin{array}{c}\Rightarrow\Rightarrow \\ \Leftarrow\Rightarrow \\ \Rightarrow\Leftarrow \\ \Leftarrow\Leftarrow\end{array}\right)\to\frac{1}{\sqrt{2}}\left(\begin{array}{cccc}1 & 0 & 0 & 0 \\ 0 & 1 & 0 & 0 \\ 0 & 0 & 0 & 1 \\ 0 & 0 & 1 & 0\end{array}\right)\left(\begin{array}{c}\Rightarrow\Rightarrow \\ \Leftarrow\Rightarrow \\ \Rightarrow\Leftarrow \\ \Leftarrow\Leftarrow\end{array}\right)\,.
\end{equation}
The CNOT gate together with the Hadamard, phase, and $\pi/8$ gates provide a fault-tolerant universal set for quantum computation.\cite{nielsenBOOK00}

\section{Fusion and read-out}
\label{fuse}

\begin{figure}
\includegraphics[width=\linewidth]{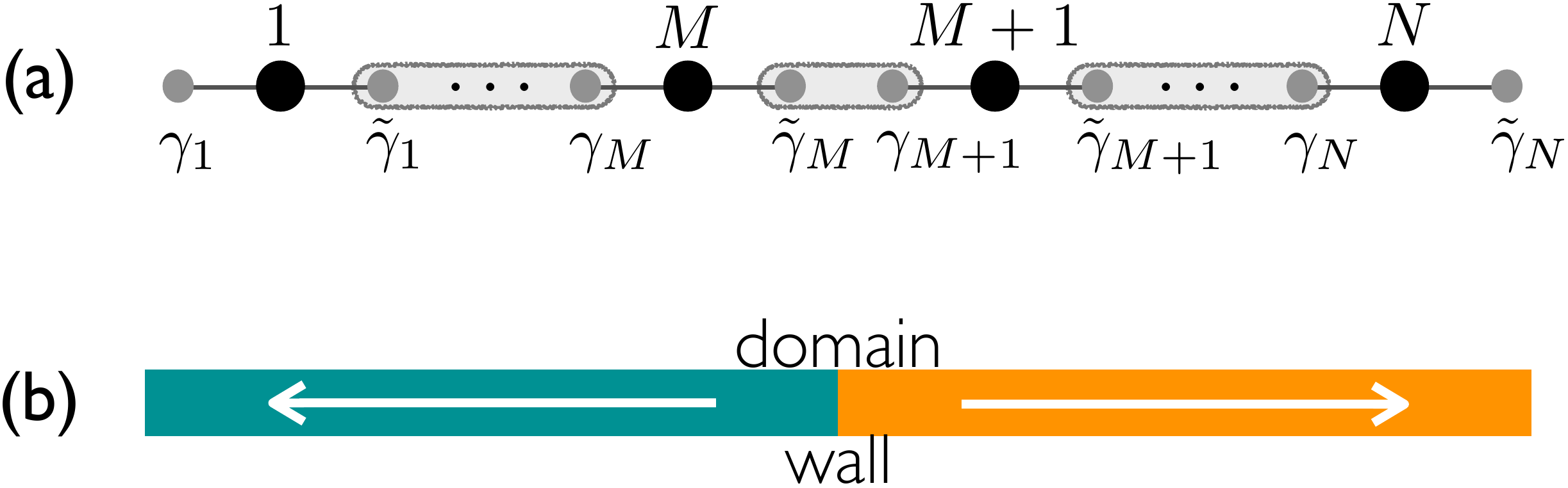}
\caption{(a) Two fused spin chains in the Majorana representation. The end Majoranas $\gamma_1$ and $\tilde{\gamma}_N$ form a fermionic zero mode $b=(\gamma_1+i\tilde{\gamma}_N)/2$, while the junction Majoranas $\tilde{\gamma}_M$ and $\gamma_{M+1}$ fuse into an ordinary fermion $a=(\tilde{\gamma}_M+i\gamma_{M+1})/2$, with energy $\epsilon=2\mathfrak{t}\cos\phi$. Here, $\mathfrak{t}$ is the exchange coupling across the $\{M,M+1\}$ junction and $\phi$ is the relative tilt of the Ising exchange interactions in the segments $1\leq i\leq M$ and $M+1\leq i\leq N$. (b) When $\phi=0$, occupying fermion state $a$ creates the spin domain wall sketched in the figure, while the domain wall corresponds to the empty fermion state when $\phi=\pi$.}
\label{fig3}
\end{figure}

Bringing two topological segments in contact along a spin chain leads to interaction between the end Majoranas. Let us model it by the following Hamiltonian:
\begin{align}
\frac{H}{J}=&-\sum_{i=1}^{M-1}\hat{\sigma}^m_i\hat{\sigma}^m_{i+1}-\sum_{i=M+1}^{N-1}\hat{\sigma}^x_i\hat{\sigma}^x_{i+1}\nonumber\\
&-\mathfrak{t}(\hat{\sigma}^x_M\hat{\sigma}^x_{M+1}+\hat{\sigma}^y_M\hat{\sigma}^y_{M+1})\,.
\label{Hf}
\end{align}
Here, two anisotropic sections $i=1\dots M$ and $i=M+1\dots N$ have Ising interactions along $\mathbf{m}=(\cos\phi,\sin\phi,0)$ and $\mathbf{x}$, respectively (recall $\hat{\sigma}_i^m=\boldsymbol{\hat{\sigma}}_i\cdot\mathbf{m}$), interacting with an isotropic (real-valued) XX coupling $\mathfrak{t}$ in the $xy$ plane at the junction, $i=\{M,M+1\}$. When $\mathfrak{t}\to0$, the two sections are disconnected, each producing two Majorana sites at the ends, as explained in the Appendix. In order to understand the low-energy properties, it is natural to switch to fermionic language, performing a phase-shifted Jordan-Wigner transformation
\begin{equation}
c_i^\dagger=\hat{\sigma}_i^+e^{i\left(\pi\sum_{j<i}\hat{n}_j-\phi\right)}
\label{cf}
\end{equation}
for $i\leq M$ and
\begin{equation}
c_i^\dagger=\hat{\sigma}_i^+e^{i\pi\sum_{j<i}\hat{n}_j}
\end{equation}
for $i>M$. Equation (\ref{Hf}) then becomes:
\begin{equation}
\frac{H}{J}=-\sum_{\langle i\rangle=1}^{N-1}(c_i^\dagger c_{i+1}+c_i^\dagger c_{i+1}^\dagger)-2\mathfrak{t} e^{i\phi}c_M^\dagger c_{M+1}+{\rm H.c.},
\label{Hcc}
\end{equation}
which can, in turn, be rewritten as [see Fig.~\ref{fig3}(a)]
\begin{equation}
\frac{H}{J}=i\sum_{\langle i\rangle=1}^{N-1}\tilde{\gamma}_i\gamma_{i+1}+i\mathfrak{t}\cos\phi\,\tilde{\gamma}_M\gamma_{M+1}+\dots\,,
\label{Hg}
\end{equation}
in terms of Majorana operators (\ref{majo}). By $\langle i\rangle$ we denote here $i\neq M$ and $\dots$ in Eq.~(\ref{Hg}) stand for terms of the form $\gamma_M\gamma_{M+1}$, $\tilde{\gamma}_M\tilde{\gamma}_{M+1}$, and $\gamma_M\tilde{\gamma}_{M+1}$, which involve the gapped states and thus have essentially no effect on the interaction of the end Majorana zero modes $\tilde{\gamma}_M$ and $\gamma_{M+1}$ when $\mathfrak{t}\ll1$. Making exchange interaction at the junction $\{M,M+1\}$ anisotropic in the $xy$ plane does not alter the essential outcome. For example, a spin exchange $-\mathfrak{t}\hat{\sigma}^{m'}_M\hat{\sigma}^{m'}_{M+1}$, where $\mathbf{m}'=(\cos\phi',\sin\phi',0)$, results in the coupling $i\mathfrak{t}\cos\phi'\cos(\phi-\phi')\tilde{\gamma}_M\gamma_{M+1}$ in fermionic language, which is reduced by a geometric factor $\cos\phi'$ and phase-shifted by $\phi'$ with respect to the isotropic XX case, Eq.~(\ref{Hg}). In particular, we see that the term $\hat{\sigma}^y_M\hat{\sigma}^y_{M+1}$ in Eq.~(\ref{Hf}) corresponds to $\phi'=\pi/2$ and thus does not couple Majoranas $\tilde{\gamma}_M$ and $\gamma_{M+1}$.

According to Hamiltonian (\ref{Hg}), the low-energy spectrum is governed by the Majorana modes $\gamma_1$, $\tilde{\gamma}_N$, $\tilde{\gamma}_M$, and $\gamma_{M+1}$. The first pair of these defines a fermionic zero mode, $b=(\gamma_1+i\tilde{\gamma}_N)/2$, while the second pair fuses into an ordinary finite-energy fermion, $a=(\tilde{\gamma}_M+i\gamma_{M+1})/2$. The total Hamiltonian in the subspace of these four Majoranas is given by
\begin{equation}
\frac{H}{J}=\mathfrak{t}\cos\phi(2a^\dagger a-1)\,.
\label{Ha}
\end{equation}
If the orientation $\phi$ of the Ising coupling in the left spin-chain segment of our system is variable and controlled by some external handle (which, e.g., defines the shape of quantum dots that provide spin states), the latter experiences backaction torque
\begin{equation}
\tau_z\equiv-\partial_\phi H=J\mathfrak{t}\sin\phi(2a^\dagger a-1)\,,
\end{equation}
in response to Hamiltonian (\ref{Ha}). If the fermion state $a$ is occupied, $\tau_z=J\mathfrak{t}\sin\phi$, while an empty state gives $\tau_z=-J\mathfrak{t}\sin\phi$. The ground-state configuration corresponding to Eq.~(\ref{Ha}) results in $\tau_z=-J{\rm sgn}(\mathfrak{t}\cos\phi)\sin\phi$. The period of the torque $\tau_z(\phi)$ in equilibrium is thus half of that for a given (either empty or occupied) eigenstate. This is fully analogous to the fractional Josephson effect in topological semiconducting wires.\cite{kitaevUFN01} Consider the system initially (in the absence of coupling $\mathfrak{t}$) prepared in the state $|i\rangle$, such that $b_1|i\rangle=b_2|i\rangle=0$, where $b_1=(\gamma_1+i\tilde{\gamma}_M)/2$ and $b_2=(\gamma_{M+1}+i\tilde{\gamma}_N)/2$ (i.e., ordinary fermions composed of end Majoranas in the disjoint topological segments). In the basis $\{1,a^\dagger,b^\dagger,a^\dagger b^\dagger\}|0\rangle$, where $a|0\rangle=b|0\rangle=0$, it is then given by $|i\rangle=(1+ia^\dagger b^\dagger)|0\rangle/\sqrt{2}$. Turning $\mathfrak{t}$ on then in Eq.~(\ref{Ha}) splits the degeneracy between states $|0\rangle$ and $a^\dagger b^\dagger|0\rangle$, resulting in opposite torques, $\mp J\mathfrak{t}\sin\phi$ (which is zero on average in $|i\rangle$).

Torque measurement of two fused qubits (each corresponding to a topological segment of the spin chain) is thus sensitive to the initial two-qubit state, providing a possible read-out mechanism. While this is technologically challenging, recent strides in quantum measurements of nanomechanical cantilevers\cite{lahayeNAT09,connellNAT10} appear encouraging to this end. In a flux qubit realization of the Ising chain,\cite{levitovCM01} on the other hand, such effective torque corresponds to a purely electronic signal.

The most straightforward way to visualize fused qubits [see Fig.~\ref{fig3}(b)] is to consider cases $\phi=0$ or $\pi$, which, according to Eq.~(\ref{Hf}), corresponds to collinear $x$-Ising segments that are weakly coupled by $-J\mathfrak{t}\hat{\sigma}^x_M\hat{\sigma}^x_{M+1}$ at the junction. (As noted above, the term $-J\mathfrak{t}\hat{\sigma}^y_M\hat{\sigma}^y_{M+1}$ has no effect on the low-energy properties, to the leading order in $\mathfrak{t}$.) The ordinary fermion formed in the junction, Eq.~(\ref{Ha}), thus corresponds to an Ising domain wall with energy $2J\mathfrak{t}$. The absence or presence of this domain wall can be revealed by simultaneous measurement of spin at the outside ends, $i=1$ and $i=N$. In the presence of some Heisenberg interaction, furthermore, the domain wall can move in response to an applied magnetic field along the chain\cite{schryerJAP74,braunPRB96} (inducing emf around the wire by the Faraday effect),\cite{kovalevSSC10} which offers alternative schemes for information read-out as well as transmission. Finally, we note that well-established spin-to-charge read-outs\cite{zakRNC10} can be applied to individual spin sites (such as quantum dots) constituting a spin-chain qubit, providing a simple mechanism for single-qubit measurements.

\section{Discussion}

External magnetic field fluctuations can in principle cause classical (i.e., flip) as well as phase errors in our logic space. Perhaps most problematically, the ferromagnetic order parameter $\langle\hat{\sigma}_x\rangle$ can couple to external magnetic fields along the $x$ direction, dephasing a single qubit in the $\{|{\Rightarrow}\rangle,|{\Leftarrow}\rangle\}$ basis or flipping it in the $\{|{+}\rangle,|{-}\rangle\}$ basis.\cite{kitaevUFN01} This issue can be mitigated by adjusting the exchange Hamiltonian of our spin chains in accordance with the structure of the dominant noise source, thus establishing the qubits in a ``decoherence-free subspace."\cite{kempePRA01} As the simplest example, in the presence of long-wavelength magnetic-field fluctuations, encoding qubits with antiferromagnetic rather than ferromagnetic spin chains should significantly reduce decoherence. Temperature, furthermore, has to be much less than the quasiparticle excitation gap scale $J$. Despite mathematical similarity of topological spin-based qubits to their superconducting Majorana cousins, the dominant quantum errors there are of a very different physical origin.\cite{Note4}

\acknowledgments

We are grateful to M. Trif and L. Trifunovic for stimulating discussions. This work was supported in part by the Alfred P. Sloan Foundation, DARPA, the NSF under Grant No. DMR-0840965 (Y.T.), and by the Swiss NSF, NCCR Nanoscience and NCCR QSIT, SOLID,  and  DARPA QuEST (D.L.). Y.T. is grateful for hospitality at the University of Basel, where much of this work has been carried out.

\appendix

\section{Generalities}
\label{gen}

Here, we review introductory material that is necessary for constructing spin-chain based topological qubits. Consider the following general anisotropic 1D spin-$1/2$ chain of $N$ sites with nearest-neighbor exchange couplings and open ends:
\begin{equation}
\frac{H}{J}=-\sum_{i=1}^{N-1}\left(\hat{\sigma}_i^x\hat{\sigma}_{i+1}^x+\alpha\hat{\sigma}_i^y\hat{\sigma}_{i+1}^y+\beta\hat{\sigma}_i^z\hat{\sigma}_{i+1}^z\right)-h\sum_{i=1}^N\hat{\sigma}_i^z\,.
\end{equation}
The Hamiltonian $H$ is normalized by the $x$-Ising exchange constant $J$, and $\alpha$, $\beta$, $h$ are dimensionless variables parametrizing the relative strength of the remaining terms. In particular, if $\alpha=\beta=0$, we have the Ising chain in transverse field $h$. $\alpha=\beta=1$ describes the Heisenberg chain and $\alpha=1$, $\beta=0$ defines the XX model. The desirable anisotropic spin Hamiltonian can be accomplished utilizing spin-orbit interactions in electrostatically coupled lateral quantum dots\cite{trifPRB07,gangadharaiahPRL08} or Wigner crystals.\cite{tserkovPRL09} Effective spin chains can, furthermore, be engineered by an array of coupled superconducting flux qubits.\cite{levitovCM01}

Performing the Jordan-Wigner transformation\cite{liebANP61}
\begin{equation}
c_i^\dagger=\hat{\sigma}_i^+e^{i\pi\sum_{j<i}\hat{n}_j}\,,
\label{c}
\end{equation}
where $\hat{\sigma}_i^\pm=(\hat{\sigma}_x\pm i\hat{\sigma}_y)/2$ and $\hat{n}_i=(\hat{\sigma}_i^z+1)/2=c_i^\dagger c_i$, defines fermionic representation for the Hamiltonian:
\begin{widetext}
\begin{equation}
\frac{H}{J}=-\sum_{i=1}^{N-1}\left[(c_i^\dagger-c_i)(c_{i+1}^\dagger+c_{i+1})-\alpha(c_i^\dagger+c_i)(c_{i+1}^\dagger-c_{i+1})\right]-\beta\sum_{i=1}^{N-1}(2\hat{n}_i-1)(2\hat{n}_{i+1}-1)-h\sum_{i=1}^N(2\hat{n}_i-1)\,.
\label{Hc}
\end{equation}
\end{widetext}
One easily verifies that $\{c_i,c_j^\dagger\}=\delta_{ij}$ and $\{c_i,c_j\}=0$, which is accomplished by the Jordan-Wigner \textit{string} $e^{i\pi\sum_{j<i}\hat{n}_j}$. We notice that the XX model gives for the first term in Eq.~(\ref{Hc}) a simple hopping Hamiltonian
\begin{equation}
-2\sum_{i=1}^{N-1}c_i^\dagger c_{i+1}+{\rm H.c.}\,,
\end{equation}
while the Heisenberg model in addition has nearest-neighbor interaction $\propto\hat{n}_i\hat{n}_{i+1}$. The field term $\propto h$ in the fermionic language enters as a chemical potential shift.

According to the form of the fermionic Hamiltonian (\ref{Hc}), it is convenient to define self-conjugate Majorana operators\cite{liebANP61,kogutRMP79,tsvelikBOOK03}
\begin{equation}
\gamma_i=c_i^\dagger+c_i\,,\,\,\,\tilde{\gamma}_i=i(c_i^\dagger-c_i)\,,\,\,\,{\rm s.t.}\,\,\,\{\gamma_I,\gamma_J\}=2\delta_{IJ}\,,
\label{majo}
\end{equation}
which can be viewed as the real and imaginary parts of the fermionic field operators, respectively. The first term of Hamiltonian (\ref{Hc}) then becomes
\begin{equation}
i\sum_{i=1}^{N-1}(\tilde{\gamma}_i\gamma_{i+1}-\alpha\gamma_i\tilde{\gamma}_{i+1})\,.
\end{equation}

\subsection{Pure Ising chain}
\label{PIC}

In the special case of a pure Ising model, $\alpha=\beta=h=0$ and thus $H/J=i\sum_{i=1}^{N-1}\tilde{\gamma}_i\gamma_{i+1}$, which means that the operators $\gamma_1$ and $\tilde{\gamma}_N$ dropped out of the Hamiltonian completely. Combining them into a new fermionic operator
\begin{equation}
b=\frac{\gamma_1+i\tilde{\gamma}_N}{2}\,,\,\,\,{\rm s.t.}\,\,\,\{b,b^\dagger\}=1\,,
\label{fb}
\end{equation}
and also pairing the remaining $2(N-1)$ Majorana operators $\gamma$'s into $(N-1)$ fermionic operators $a_i=(\tilde{\gamma}_i+i\gamma_{i+1})/2$, such that $H=\hat{a}^\dagger\hat{A}\hat{a}$, with some Hermitian $(N-1)\times(N-1)$ matrix $\hat{A}$, we conclude that all eigenstates corresponding to $\hat{A}$ acquire double degeneracy due to $b$. Namely, for all eigenstates $|0\rangle_a$ such that $H|0\rangle_a=\epsilon_a|0\rangle_a$ and $b|0\rangle_a=0$, there is also an orthogonal degenerate eigenstate $|1\rangle_a=b^\dagger|0\rangle_a$. Most importantly for us, this concerns also the degenerate ground states $|0\rangle$ and $|1\rangle$.

The end \textit{Majoranas} that dropped out of the Hamiltonian, in the case of a pure Ising model, give rise to double degeneracy of the ground state. This is of course obvious in the original spin representation, according to which we trivially have two ground states:
\begin{equation}
|{\Rightarrow}\rangle=|{\rightarrow\rightarrow\dots\rightarrow}\rangle\,\,\,{\rm and}\,\,\,
|{\Leftarrow}\rangle=|{\leftarrow\leftarrow\dots\leftarrow}\rangle\,,
\label{gs}
\end{equation}
where
\begin{equation}
|{\rightarrow}\rangle=\frac{1}{\sqrt{2}}\left(\begin{array}{c}1\\1\end{array}\right)\,\,\,{\rm and}\,\,\,
|{\leftarrow}\rangle=\frac{1}{\sqrt{2}}\left(\begin{array}{c}1\\-1\end{array}\right)
\end{equation}
are the single-spin eigenstates of $\hat{\sigma}^x$, and we assumed $J>0$, to be specific.

In addition to the time-reversal symmetry, the ground states (\ref{gs}) spontaneously break the following $\mathbb{Z}_2$ symmetry:
\begin{equation}
\mathcal{P}=\prod_{i=1}^N\hat{\sigma}^z_i\,,
\label{Ps}
\end{equation}
which flips $|{\Rightarrow}\rangle$ into $|{\Leftarrow}\rangle$ and vice versa (flipping the associated magnetic order parameter $\langle\hat{\sigma}^x\rangle$). In the fermionic language,
\begin{equation}
\mathcal{P}=\prod_{i=1}^N(2\hat{n}_i-1)=(-1)^{\sum_{i=1}^N(\hat{n}_i-1)}
\label{P}
\end{equation}
is the parity operator, with eigenstates $\pm1$ corresponding respectively to the even/odd number of fermionic holes present in the chain. Since the ground states $|0\rangle$ and $|1\rangle$ have well-defined parity, they must correspond to the symmetric and antisymmetric combinations of $|{\Rightarrow}\rangle$ and $|{\Leftarrow}\rangle$. Specifically,
\begin{equation}
|0\rangle=\frac{|{\Rightarrow}\rangle+|{\Leftarrow}\rangle}{\sqrt{2}}\,,\,\,\,|1\rangle=\frac{|{\Rightarrow}\rangle-|{\Leftarrow}\rangle}{\sqrt{2}}
\label{01}
\end{equation}
if $N$ is even and vice versa if $N$ is odd. In the spin language, the Majorana fermion operator $b$ (such that $b|1\rangle=|0\rangle$ and $b|0\rangle=0$) is
\begin{equation}
b=\frac{\hat{\sigma}_1^x-\hat{\sigma}^x_NZ}{2}\,,
\end{equation}
where $Z=\prod_{i=1}^N(-\sigma_i^z)$ is the \textit{string} operator. Note that since $b$ is a nonlocal operator because of the string $Z$, the degenerate ground states $|0\rangle$ and $|1\rangle$ cannot be interpreted in terms of local spin states. The $\mathbb{Z}_2$ symmetry (\ref{Ps}), furthermore, stabilizes the degeneracy against local weak parity-conserving perturbations.\cite{kitaevUFN01}

It is important to point out a key difference between the aforementioned degeneracy in a spin chain vs a topological superconducting wire.\cite{kitaevUFN01} While the systems become mathematically equivalent in 1D with open boundary conditions, it is not so once we switch to periodic boundary conditions.\cite{liebANP61} In particular, the Ising chain preserves degeneracy (which, according to the time-reversal symmetry, is unaffected by additional Heisenberg and/or Dzyaloshinsky-Moriya interactions, as long as some magnetic ordering remains), while the periodic boundary conditions in general lift the degeneracy of Majorana end states in the superconducting wires. This means that there is nothing fundamentally special about the end regions in our spin chains, and the Majorana-fermion analogy with topological superconducting wires is only fulfilled for open boundary conditions.

While both open and cyclic spin chains allow for the ground-state degeneracy, open chains are expected to be more robust to higher-order global $|{\Rightarrow}{\rangle}\leftrightarrow|{\Leftarrow}{\rangle}$ flipping processes than their closed counterparts. A closed chain can be flipped by a virtual local single-spin flip event (at an arbitrary location) that costs energy $\sim2J$, which then produces two domain walls that propagate around the chain and annihilate each other (thus returning the extra energy) after flipping the entire chain. In an open spin chain, on the other hand, the domain walls would need to be nucleated either at the ends or exactly at the center of the chain, or else the extra energy $~2J$ would be returned at the two ends (one half of it at a time) at different times, thus requiring an additional virtual process. This observation reveals a topological character of our spin-chain logic-state subspace, which is distinct from the topological nature of the Majorana end states in superconducting wires.

\subsection{Ising chain in a transverse magnetic field}

Let us look more closely at the problem of Ising chain in a transverse magnetic field:
\begin{equation}
\frac{H}{J}=-\sum_{i=1}^{N-1}\hat{\sigma}_i^x\hat{\sigma}_{i+1}^x-h\sum_{i=1}^N\hat{\sigma}_i^z\,.
\label{Hi}
\end{equation}
While the time-reversal symmetry is broken, $\mathcal{P}$ still generates a $\mathbb{Z}_2$ symmetry of Hamiltonian (\ref{Hi}). In the thermodynamic limit, the ground state remains degenerate for $|h|<1$, while a paramagnetic phase with a unique ground state sets in for $|h|>1$. $|h|=1$ is thus the critical point for a 2nd-order quantum phase transition.\cite{sachdevBOOK99}

Diagonalizing Hamiltonian (\ref{Hi}) is most convenient in the fermionic representation (\ref{Hc}),
\begin{equation}
\frac{H}{J}=-\sum_{i=1}^{N-1}(c_i^\dagger-c_i)(c_{i+1}^\dagger+c_{i+1})-h\sum_{i=1}^N(2\hat{n}_i-1)\,,
\label{Hif}
\end{equation}
which is easily accomplished by the Bogolyubov transformation for the momentum eigenstates:\cite{sachdevBOOK99}
\begin{equation}
c_k=\frac{1}{\sqrt{N}}\sum_je^{-ikj}c_j\,,\,\,\,\gamma_k=\cos\theta_k\,c_k+i\sin\theta_k\,c_{-k}^\dagger\,,
\end{equation}
where $k=2\pi(n/N)\in(-\pi,\pi]$ and
\begin{equation}
\theta_k=\frac{1}{2}\tan^{-1}\frac{\sin k}{\cos k+h}\,.
\end{equation}
Here, for simplicity, we are imposing periodic boundary conditions for Hamiltonian (\ref{Hif}), which does not affect bulk properties.\cite{liebANP61} In terms of these Bogolyubov quasiparticles, the Hamiltonian (in the thermodynamic limit) becomes (dropping an unimportant overall constant)\cite{Note5}
\begin{equation}
H=\sum_k\epsilon_k\gamma_k^\dagger\gamma_k\,,\,\,\,\epsilon_k=2J\sqrt{1+h^2+2h\cos k}\,.
\end{equation}
The gap in the excitation spectrum closes at $|h|=1$, at which point we have a phase transition from the ordered phase with degenerate ground state (for $|h|<1$),\cite{liebANP61} to an ordinary quantum paramagnet phase with a nondegenerate ground state (for $|h|>1$). Introducing exchange interactions along the $z$ axis leads to nearest-neighbor interactions in the fermionic language, Eq.~(\ref{Hc}). This needs to be avoided as strong interactions can suppress the superconducting gap that stabilizes ground-state degeneracy.\cite{gangadharaiahCM11}

In the case of a vanishing field $h$, the degeneracy of the ground state is exact, irrespective of the length of the chain. This means that the end spin Majoranas are noninteracting or, in other words, they have a vanishing correlation length $\xi$. A small finite magnetic field (and/or more general exchange interactions) gives rise to a finite length $\xi$, while still preserving the topological nature of the ground-state doublet of the wire whose length $L\gg\xi$.\cite{liebANP61} Interaction between the end Majoranas is then generally given by
\begin{equation}
H'=i\mathfrak{t}\gamma_1\tilde{\gamma}_N\,,
\end{equation}
where $\mathfrak{t}\sim e^{-\xi/L}$ is the tunneling amplitude for a domain wall to propagate along the length of the chain, zipping $|{\Rightarrow}\rangle$ into $|{\Leftarrow}\rangle$ and vice versa.\cite{kitaevUFN01} In terms of the complex fermions (\ref{fb}), this Hamiltonian is given by $2\mathfrak{t} b^\dagger b$, such that the lowest-energy splitting is $2\mathfrak{t}$ (while the other excitations are gapped approximately by the domain-wall energy of $2J$).

Note that the sign of $J$ (i.e., the ferromagnetic, $J>0$, vs antiferromagnetic, $J<0$, case) is of no consequence, since it can be flipped in Hamiltonian (\ref{Hi}) by a unitary transformation that amounts to a $\pi$ rotation around the $z$ axis (thus $\hat{\sigma}^z\to\hat{\sigma}^z$ and $\hat{\sigma}^{x,y}\to-\hat{\sigma}^{x,y}$) for all odd sites.\cite{Note6}

Finally, performing a global $\phi$ rotation around the $z$ axis, we obtain the following Hamiltonian:
\begin{equation}
\frac{H}{J}=-\sum_{i=1}^{N-1}\hat{\sigma}_i^m\hat{\sigma}_{i+1}^m-h\sum_{i=1}^N\hat{\sigma}_i^z\,,
\end{equation}
where $\hat{\sigma}_i^m=\boldsymbol{\hat{\sigma}}_i\cdot\mathbf{m}$ and $\mathbf{m}=(\cos\phi,\sin\phi,0)$. In the fermionic language, Eq.~(\ref{c}), this rotation generates a global gauge transformation: $c_i\to c_ie^{i\phi}$. The Ising interaction thus becomes
\begin{align}
\hat{\sigma}_i^m\hat{\sigma}_{i+1}^m&=(c_i^\dagger e^{-i\phi}-c_ie^{i\phi})(c_{i+1}^\dagger e^{-i\phi}+c_{i+1}e^{i\phi})\nonumber\\
&=c_i^\dagger c_{i+1}+c_i^\dagger c_{i+1}^\dagger e^{-2i\phi}+{\rm H.c.}\,,
\end{align}
so that a $\phi$ rotation in spin space is equivalent to the $2\phi$ phase change of the superconducting order parameter, $\Delta\to\Delta e^{-2i\phi}$, in the fermionic language. In particular, rotating the Ising axis by $\pi/2$ [see Fig.~\ref{fig2}(b)] is equivalent to changing the sign of the superconducting gap, which explains why the geometric manipulations in Fig.~\ref{fig2}(b) mimic Majorana braiding (and the associated phase gate) in the superconducting wires.\cite{aliceaNATP11}

\end{document}